# Recent Progress in the Definition of Thermodynamic Entropy[1]

**Enzo Zanchini**

Università di Bologna, Department of Industrial Engineering, Viale Risorgimento 2, 40136 Bologna, Italy
enzo.zanchini@unibo.it

and

**Gian Paolo Beretta**

Università di Brescia, Department of Mechanical and Industrial Engineering, Via Branze 38, 25123 Brescia, Italy
gianpaolo.beretta@unibs.it

**Abstract:** The principal methods for the definition of thermodynamic entropy are discussed with special reference to those developed by Carathéodory, the Keenan School, Lieb and Yngvason, and the present authors. An improvement of the latter method is then presented. Seven basic axioms are employed: three *Postulates*, which are considered as having a quite general validity, and four *Assumptions*, which identify the domains of validity of the definitions of energy (Assumption 1) and entropy (Assumptions 2, 3, 4). The domain of validity of the present definition of entropy is not restricted to stable equilibrium states. For collections of simple systems, it coincides with that of the proof of existence and uniqueness of an entropy function which characterizes the relation of adiabatic accessibility proposed by Lieb and Yngvason. However, our treatment does not require the formation of *scaled copies* so that it applies not only to collections of simple systems, but also to systems contained in electric or magnetic fields and to small and few-particle systems.

## 1. Introduction

From the origins of classical thermodynamics to the present time, several methods for the definitions of thermodynamic temperature and thermodynamic entropy have been developed. If we exclude the treatments based on statistical mechanics and those which postulate directly the existence and additivity of entropy, as well as the structure of the fundamental relations [1], most of the methods proposed in the literature, up to the most recent contributions, can be divided in three main categories: classical methods, Carathéodory-derived methods, Keenan-school methods.

Classical methods start with the Zeroth-Law of thermodynamics (transitivity of mutual thermal equilibrium) [2, 3] and the definition of empirical temperature, then define energy by a suitable statement of the First Law, and finally define thermodynamic temperature and entropy through the Kelvin-Planck statement of the Second Law [4], namely, it is impossible to construct an engine which,

---

operating in a cycle, produces no effect except the lifting of a weight and the cooling of a thermal reservoir.

In their original formulation, classical methods had a logical loop in the definition of energy. In fact, the First Law was stated as follows: in a cycle, the work done by a system is equal to the heat received by the system,

$$Q = W. \tag{1}$$

The energy difference between state $A_2$ and state $A_1$ of a system $A$ was defined as the value of $Q - W$ in any process for $A$ from $A_1$ to $A_2$. Clearly, this definition is affected by a logical circularity, because it is impossible to define heat without a previous definition of energy.

The circularity of Eq. (1) was understood and resolved in 1909 by Carathéodory [5] who defined an adiabatic process without employing the concept of heat as follows. A vessel is called *adiabatic* if the state of the system inside it does not change when the bodies present outside the vessel are modified, provided that the vessel remains at rest and retains its original shape and volume. A process such that the system is contained within an adiabatic vessel at every instant of time is called an *adiabatic process*. Carathéodory stated the First Law as follows: *the work performed by a system in any adiabatic process depends only on the end states of the system.*

Among the best treatments of thermodynamics by the classical method, we can cite, for instance, those by Fermi [6] and by Zemansky [3]. In these treatments, Carathéodory's statement of the First Law is adopted.

In his celebrated paper [5], Carathéodory also proposed a new statement of the Second Law and developed a completely new method for the definitions of thermodynamic temperature and entropy. The treatment refers to simple systems, stable equilibrium states, and quasistatic processes, *i.e.*, processes in which the system evolves along a sequence of neighboring stable equilibrium states. A simple system is defined by Carathéodory as a system such that

(a) its stable equilibrium states are determined uniquely by $n + 1$ coordinates, $\xi_0, x_1, \ldots, x_n$, where $x_1, \ldots, x_n$ are deformation coordinates, *i.e.*, coordinates which determine the external shape of the system, while $\xi_0$ is not a deformation coordinate;
(b) in every reversible quasistatic process, the work performed by the system is given by

$$\delta W = p_1 \, d x_1 + \ldots + p_n \, d x_n \tag{2}$$

where $p_1, \ldots, p_n$ are functions of $\xi_0, x_1, \ldots, x_n$; and
(c) the (internal) energy $U$ of the system, which is defined via the First Law, is additive, *i.e.*, equals the sum of the energies of its subsystems.

Carathéodory stated the Second Law (Axiom II) as follows: *in every arbitrarily close neighborhood of a given initial state there exist states that cannot be reached by adiabatic processes.* By employing a mathematical theorem on Pfaffian equations, he proved that, on account of the Second Law, there exists a pair of properties, $M(\xi_0, x_1, \ldots, x_n)$ and $x_0(\xi_0, x_1, \ldots, x_n)$ such that for every quasistatic process

$$dU + \delta W = M \, d x_0. \tag{3}$$

Through other assumptions on the conditions for mutual stable equilibrium, which include the Zeroth Law (transitivity of mutual stable equilibrium), Carathéodory proved that there exists a function $\tau(x_0, x_1, \ldots, x_n)$ called temperature such that if two systems $A$ and $B$ are in mutual stable equilibrium they have the same temperature. Moreover, by applying the additivity of energy, he proved that there exists a function $f(\tau)$, identical for all systems, such that

$$M = f(\tau)\,\alpha(x_0) \ , \tag{4}$$

where $\alpha(\cdot)$ is another function that varies from system to system.

He then defined thermodynamic temperature $T$ and entropy $S$ respectively as

$$T = c\,f(\tau), \quad S - S_{ref} = \int_{x_{0,ref}}^{x_0} \frac{\alpha(x_0')}{c}\,\mathrm{d}x_0' \ , \tag{5}$$

where $c$ is an arbitrary constant and $S_{ref}$ is an arbitrary value assigned to the reference state with $x_0 = x_{0,ref}$. Finally, he rewrote Eq. (3) in the form

$$\mathrm{d}U + \delta W = T\,\mathrm{d}S \ , \tag{6}$$

which, through Equation (2), yields the Gibbs relation for the stable equilibrium states of a simple system.

Carathéodory's method for the definition of entropy has been the point of origin for several efforts along a similar line of thought. Some authors tried to simplify the treatment and make it less abstract [7-9], while others tried to improve the logical rigor and completeness [10]. Several references on the Carathéodory-derived methods for the definition of entropy are quoted and discussed in a recent paper by Lieb and Yngvason [11].

An alternative method for the treatment of the foundations of thermodynamics was introduced by Keenan [12] and developed by Hatsopoulos and Keenan [13] and by Gyftopoulos and Beretta [14]: it will be called the Keenan-school method. An advantage of this method, with respect to that of Carathéodory, is that the treatment does not employ the concepts of simple system and of quasistatic process so that it applies also to systems contained in electric and magnetic fields and, at least potentially, also to nonequilibrium states. Another step forward introduced by the Keenan school and developed particularly in Ref. [14] is the statement of a broad set of operational definitions of the basic concepts of thermodynamics such as those of system, property, state, isolated system, environment of a system, and thermal reservoir. The concept of adiabatic process introduced by Carathéodory is replaced in Ref. [14] by the simpler and less restrictive concept of *weight process*: a process of a system $A$ such that the only net effect in the environment of $A$ is a purely mechanical effect such as, for instance, the raising of a weight in a gravitational field or the displacement of an electric charge in a uniform electrostatic field. The main difference between an *adiabatic process* and a *weight process* is that while the former implies a constraint on the whole time evolution of system $A$ and its environment, the latter allows any kind of interaction during the process and sets a constraint only on the end states of the environment of $A$.

The *First Law* is stated as follows. *Any two states of a system can be the end states of a weight process. Moreover, the work performed by a system in any weight process depends only on the end states of the system* [14].

The first part of the statement makes explicit a condition which is almost always employed implicitly when energy is defined through the Carathéodory statement of the First Law. However, the general validity of this condition is questioned in Ref. [11]. Indeed, it is possible to release this condition as shown in Ref. [15].

A *parameter* of a system *A* is defined in Ref. [14] as a physical quantity determined by a real number, which describes an overall effect on *A* of bodies in the environment of *A*, such as the volume *V* of a container or the gravitational potential. An *equilibrium* state of a system *A* is defined as a stationary state of *A* which can be reproduced as a stationary state of an isolated system. An equilibrium state of *A* is called a *stable equilibrium state* if it cannot be modified in a process which leaves unchanged both the parameters of *A* and the state of the environment of *A*. The following statement of the *Second Law* is employed [14]. *Among all the states of a system that have a given value E of the energy and are compatible with a given set of values **n** of the amounts of constituents and **β** of the parameters, there exists one and only one stable equilibrium state. Moreover, starting from any state of a system, it is always possible to reach a stable equilibrium state with arbitrarily specified values of amounts of constituents and parameters by means of a reversible weight process.*

The second part of this statement of the *Second Law* is very demanding and should be revised and/or clarified. Indeed, an arbitrary change in composition of a system *A* (such as one which requires destruction or creation of matter) cannot be obtained by a weight process for *A* (without the corresponding creation or destruction of antimatter).

The definition of entropy is given by employing two auxiliary quantities, called *generalized adiabatic availability* and *generalized available energy*. This method for the definition of entropy has the advantage of emphasizing from the beginning the relation between entropy and the maximum work obtainable in a weight process with a given initial state but has the disadvantage of making the treatment longer and more complex.

In recent years, two novel approaches to the definition of entropy have been introduced by Lieb and Yngvason [11, 16, 17] and by the present authors [18, 19].

The definition of entropy proposed by Lieb and Yngvason is based on the concept of *adiabatic accessibility*. A state *Y* is said to be *adiabatically accessible* from a state *X*, in symbolic form $X \prec Y$, if it is possible to change the state from *X* to *Y* by means of an interaction with some device and a weight in such a way that the device returns to its initial state at the end of the process whereas the weight may have changed its position in a gravitational field.

Note that the concept of *adiabatic accessibility* defined above coincides with that of *accessibility by means of a weight process* as defined in Ref. [14] and not with the adiabatic accessibility considered by Carathéodory.

If $X \prec Y$ but *X* is not adiabatically accessible from *Y*, the symbol $X \prec\prec Y$ is employed. If both $X \prec Y$ and $Y \prec X$, then *X* is said to be equivalent to *Y*, $X \sim Y$. Another key concept in the Lieb-Yngvason approach is that of *scaled copy* of a system. Let *t* be any positive real number, $\Gamma$ a system, and *X* a state of $\Gamma$. The *t-scaled copy* $\Gamma^{(t)}$ of $\Gamma$ is a system such that to every state *X* of $\Gamma$ there corresponds a state *tX* of $\Gamma^{(t)}$ where the values of all the extensive properties (volume, energy, mole numbers …) are equal to *t* times the values they have in state *X*.

The following axioms on the order relation $\prec$ are postulated:
(A1) *Reflexivity*. $X \sim X$.
(A2) *Transitivity*. $X \prec Y$ and $Y \prec Z$ implies $X \prec Z$.

(A3) *Consistency.* $X \prec Y$ and $X' \prec Y'$ implies $(X, X') \prec (Y, Y')$, where the pairs of states in brackets are states of the composite system $(\Gamma, \Gamma')$.

(A4) *Scaling invariance.* If $X \prec Y$, then $tX \prec tY$ for every $t > 0$.

(A5) *Splitting and recombination.* For $0 < t < 1$, $X \sim (tX, (1-t)X)$.

(A6) *Stability.* If, for some pairs of states $X$ and $Y$, $(X, \varepsilon Z_0) \prec (Y, \varepsilon Z_1)$ holds for a sequence of $\varepsilon$'s tending to zero and some states $Z_0$ and $Z_1$, then $X \prec Y$.

While Axioms (A1), (A2), and (A3) describe properties that one expects as natural for the adiabatic accessibility relation among states, the other axioms, which refer to *scaled copies* of a system, limit in effect the validity of the treatment presented in Refs. [11, 16] to systems which can be considered as the union of simple systems, each in a stable equilibrium state. For instance, scaled copies cannot be formed for a dielectric system in an electrostatic field nor for a small system with non-negligible, non-local effects and/or near-wall rarefaction or capillarity effects.

Lieb and Yngvason consider a set of stable equilibrium states, which they call *state space* and denote by $\Gamma$, where the following assumption holds:

*Comparison hypothesis.* Any two states $X$ and $Y$ in $\Gamma$ are comparable, i.e., either $X \prec Y$ or $Y \prec X$ or both.

They prove that in a state space $\Gamma$ such that the *comparison hypothesis* holds for $\Gamma$ and for every scaled copy of $\Gamma$, axioms (A1) ÷ (A6) imply the *entropy principle* stated as follows.

There is a real-valued function on all the stable equilibrium states of all systems (including compound systems), called *entropy* and denoted by $S$ such that

(a) when $X$ and $Y$ are comparable states, then $X \prec Y$ if and only if $S(X) \leq S(Y)$;

(b) if $X$ and $Y$ are states of some (possibly different) systems 1 and 2 and if $(X, Y)$ denotes the corresponding state in the state space $\Gamma_1 \times \Gamma_2$ of the composite of the two systems, then the entropy is additive for these states, namely $S((X, Y)) = S(X) + S(Y)$;

(c) $S$ is extensive, i.e., for each state $X$ in state space $\Gamma$ and every integer $t > 0$, we can write $S((tX)) = tS(X)$ where state $(tX)$ is the scaled copy of $X$ belonging to the $t$-times scaled copy of $\Gamma$, namely $\Gamma^{(t)} = \Gamma \times \cdots_{t\text{ times}} \times \Gamma$.

For a given state space $\Gamma$, the entropy $S$ can be determined by the following procedure. Let us consider two states in $\Gamma$, $X_0$ and $X_1$, such that $X_0 \prec\prec X_1$ and, therefore, on account of the entropy principle, $S(X_0) < S(X_1)$. Let $X$ be a state in $\Gamma$, such that $X_0 \prec X \prec X_1$. Then, it is possible to prove that there exists a unique real number $\lambda$ between 0 and 1 such that $X \sim ((1-\lambda)X_0, \lambda X_1)$. Once $\lambda$ has been determined, one has that

$$S(X) = (1-\lambda)S(X_0) + \lambda S(X_1) \ . \tag{7}$$

By the procedure explained above and Eq. (7) one determines an entropy function on $\Gamma$ which is unique for every choice of the reference states $X_0$ and $X_1$ or, equivalently, which is determined up to an arbitrary linear transformation: $S(X) \to aS(X) + B$ with $a > 0$. The entropy functions established on the single state space will not, in general, fulfill the additivity for different systems. However, it is possible to prove the following conclusion [16].

Assume that the *comparison hypothesis* holds for every multiple scaled copy of $\Gamma$ of the type $\Gamma^{(t_1)} \times \cdots \times \Gamma^{(t_N)}$ and let $S_\Gamma$ be an entropy function defined on $\Gamma$. Then there exist constants $a_\Gamma$ and $B(\Gamma)$ such that the function $S$ defined for all the stable equilibrium states of all state spaces $\Gamma^{(t_1)} \times \cdots \times \Gamma^{(t_N)}$ by the relation

$$S(X) = a_\Gamma S_\Gamma(X) + B(\Gamma) \tag{8}$$

for $X \in \Gamma$, satisfies additivity, i.e., $S((X, Y)) = S(X) + S(Y))$, extensivity, i.e., $S(tX) = t S(X)$, and is such that whenever $X$ and $Y$ are in the same state space, then $X \prec Y$ if and only if $S(X) \leq S(Y)$.

Another result obtained by Lieb and Yngvason in Refs. [11, 16] is to prove that the *comparison hypothesis* can be deduced, for simple systems and their scaled products, as a consequence of nine other axioms, which also imply additional structure of the state spaces and lead, e.g., to the concepts of pressure and temperature and to the concavity of entropy. These results will not be discussed here.

While Refs. [11, 16] refer exclusively to stable equilibrium states, very recently Lieb and Yngvason [17] presented an extension of the definition of entropy to a *state space* $\hat{\Gamma}$ which contains also nonequilibrium states (but not all nonequilibrium states) and is such that the *comparison hypothesis* can be considered as valid in $\hat{\Gamma}$. The authors denote by $\Gamma$ the subset of the stable equilibrium states and assume that in $\Gamma$ the relation $\prec$ fulfils Axioms (A1) ÷ (A6), moreover,

(N1) in the rest of $\hat{\Gamma}$ the relation $\prec$ fulfils Axioms (A1), (A2), (A3), and (A6); and

(N2) for every $X \in \hat{\Gamma}$ there exist $X'$, $X'' \in \Gamma$ such that $X' \prec X \prec X''$.

They then prove that the validity of Axioms (A1) ÷ (A6) in $\Gamma$, together with the validity of Axioms (N1), (N2) and the *comparison hypothesis* in the new state space $\hat{\Gamma}$ implies the existence and uniqueness (up to a linear transformation) of an entropy function $\hat{S}$ in $\hat{\Gamma}$, which extends $S$ (i.e., coincides with $S$ in the subset $\Gamma$ of the stable equilibrium states) and is such that $X \prec Y$ implies $\hat{S}(X) \leq \hat{S}(Y)$ for any two states $X$ and $Y$ in $\hat{\Gamma}$.

The contribution provided by Lieb and Yngvason [11, 16, 17] is an interesting and rigorous analysis of the conditions required for the definition of entropy in a set of states which can also contain nonequilibrium states. However, the definition of entropy along the Lieb-Yngvason method does not suggest a simple direct procedure to measure the entropy difference between any pair of equilibrium or nonequilibrium states. Moreover the method is not applicable, in its present form, to systems contained in external electric or magnetic fields nor to small scale and few particle systems, i.e., systems that do not allow the creation of scaled copies even in their stable equilibrium states.

A different approach to the definition of entropy, inspired by the Keenan-school method but independent of it, was recently developed by the present authors [18,19]. This approach is based on a set of operational definitions of the basic concepts (system, property, state, isolated system, environment of a system, weight process, equilibrium state, stable equilibrium state) very similar to that presented in Ref. [14] and on the same statement of the *First Law*. On the other hand, only the first part of the *Second Law* statement of Ref. [14] is retained. The second part is replaced by a separate assumption, which *refers only to closed systems* and *is not claimed to have a fully general validity*. The definition of entropy is not based any more on the concepts of generalized adiabatic availability and of generalized available energy. It is instead obtained directly from the basic assumptions and from the definition of energy. Moreover it is stated for *closed systems* and *extended to open systems* [19] only at a later stage of the development. This sharp separation between the treatments of closed and open systems singles out an *important conceptual dissimilarity* between the meaning of the energy (and entropy) difference between two states of a closed system and that of the energy (and entropy) difference between two states of an open system. Indeed, for a closed system, the energy difference between two states depends only on the unit fixed (once and for all) for energy, and the entropy difference depends only on the units fixed (once and for all) for energy and for thermodynamic temperature. In contrast, for an open system, the energy (and entropy) difference between two states depends also on the choices of the reference states for pure substances and on the energy (and entropy) values assigned to the pure substances in such reference states.

In analogy with Ref. [14], the concepts of *separable systems* and of *systems uncorrelated from each other* are introduced. Separability is a requirement for the definition of energy: if a system *A* is a source of long-range force fields for a system *B*, the energy of *A* (or of *B*) cannot be defined. Consider, for instance, two rigid bodies *A* and *B* in a gravitational field produced by the systems themselves: while the potential energy of the composite system *AB* can be defined, that of *A* or of *B* cannot. Non-correlation is a requirement imposed for the definition of the thermodynamic entropy, relevant for microscopic or mesoscopic systems, which can exhibit quantum-mechanical correlations with each other.

In this paper, a further development of the method for the definitions of energy and entropy proposed in Refs. [18, 19] is presented, with the following improvements. Each of the axioms employed in the treatment is placed in one of two different categories: *Postulates*, which are statements with a declared fully general validity; *Assumptions*, which are not claimed to hold necessarily for all systems and states and as such determine the *domain of validity* of the definitions of energy and of entropy derived from them. In agreement with Lieb and Yngvason [11, 16, 17], the statement that any two states of a closed system can be the end states of a weight process (i.e., the *comparison hypothesis*) is not considered as having necessarily a general validity; this statement is not employed in the present treatment. The definition of the energy difference between any pair of states of a closed system is obtained through the following axioms:

**Assumption 1.** *Any pair of states of a closed system A, such that the system is separable from its environment in both states, can be interconnected by a weight polygonal for A.*

The concept of *weight polygonal*, already employed in Ref. [15], is defined in Section 3.

**Postulate 1.** *The works done by a system in any two weight polygonals between the same initial and final states are identical.*

The definition of entropy is based on two additional Postulates and three additional Assumptions.

**Postulate 2.** *Among all the states of a closed system A such that the constituents of A are contained in a given set of regions of space, there is a stable equilibrium state for every value of the energy $E^A$.* **Assumption 2.** *Starting from any state in which the system is separable from its environment, a closed system A can be changed to a stable equilibrium state with the same energy by means of a zero work weight process for A in which the regions of space occupied by the constituents of A have no net changes.*

**Postulate 3**. *There exist systems, called normal systems, whose energy has no upper bound. Starting from any state in which the system is separable from its environment, a normal system A can be changed to a non-equilibrium state with arbitrarily higher energy by means of a weight process for A in which the regions of space occupied by the constituents of A have no net changes.*

**Assumption 3**. *There exist systems, called thermal reservoirs, which fulfill the following conditions:*
*(a) the regions of space occupied by the constituents are fixed;*
*(b) if R is a thermal reservoir in an arbitrary stable equilibrium state $R_1$ and $R^d$ is an identical copy of R also in arbitrary stable equilbrium state $R^d_2$, not necessarily equal to $R_1$, then R and $R^d$ are in mutual stable equilibrium.*

**Assumption 4**. *Every pair of states $(A_1, A_2)$ of a closed system A, such that A is separable and uncorrelated from its environment in both states, can be interconnected by a reversible standard weight process for AR, where R is an arbitrarily chosen thermal reservoir initially in an arbitrarily chosen stable equilibrium state.*

A *standard weight process* for *AR* is a weight process for the composite of systems *A* and *R* in which the end states of *R* are stable equilibrium states [18, 19] (see definition in Section 4).

It is shown in Section 5 that, if one disregards the conditions of separability and non-correlation (not discussed in Refs. [11, 16, 17]), one can establish a relation between the domain of validity of the present definition of entropy and that of the definition developed by Lieb and Yngvason. Namely, for systems which allow the creation of *scaled copies* of their state space when their subsystems are in stable equilibrium states (i.e., collections of *simple systems*), the domain of validity of the present definition of entropy coincides with the domain of the states in which Lieb and Yngvason prove the existence and uniqueness of an entropy function which characterizes the relation of adiabatic comparability [17]. However, the domain of validity of the present definition of entropy extends also to systems which do not allow the creation of *scaled copies* even in their stable equilibrium states such as systems contained in electric or magnetic fields, systems with small number of particles for which wall rarefaction effects cannot be neglected, or multi-phase systems for which capillary and interfacial effects are important.

**2. Summary of basic definitions**

We briefly recall here some definitions of the basic concepts of thermodynamics employed in our treatment. A more complete and more detailed set of operational basic definitions can be found in Refs. [18, 19].

With the term *system* we mean a set of material particles, of one or more kinds, such that, at each instant of time, the particles of each kind are contained within a given region of space. If the external surface of the union of the regions of space which contain the particles of the system cannot be crossed by any material particle, the system is called *closed*.

Any system is endowed with a set of reproducible measurement procedures such that each procedure, if applied at an instant of time *t*, yields a result which is independent of the previous time evolution of the system. Each procedure of this kind defines a *property* of the system. The set of all the values of the properties of a system, at a given instant of time, defines the *state* of the system at that instant.

A system can be in contact with other matter or surrounded by empty space. Moreover, force fields due to external matter can act in the region of space occupied by the system. If, at an instant of time, all the particles of the system are removed from the respective regions of space and brought far away, but a force field is still present in the region of space (previously) occupied by the system, then this force field is called an *external force field*. An external force field can be either gravitational, or electric or magnetic, or a superposition of the three.

Consider the union $\Sigma$ of all the regions of space spanned by a system during its entire time evolution. If no other material particles, except those of the system, are present in $\Sigma$ or touch the boundary of $\Sigma$, and if the external force field in $\Sigma$ is either vanishing or stationary, then we say that the system is *isolated*. Suppose that an isolated system $I$ can be divided into two subsystems, $A$ and $B$. Then, we can say that $B$ is the *environment* of $A$ and vice-versa.

If, at a given instant of time, two systems $A$ and $B$ are such that the force field produced by $B$ is vanishing in the region of space occupied by $A$ and vice-versa, then we say that $A$ and $B$ are *separable* at that instant. The energy of a system $A$ is defined only for the states of $A$ such that $A$ is separable from its environment.

If, at a given instant of time, two systems $A$ and $B$ are such that the outcomes of the measurements performed on $B$ are statistically independent of those of the measurements performed on $A$, and vice-versa, we say that $A$ and $B$ are *uncorrelated from each other* at that instant. The entropy of a system $A$ is defined only for the states of $A$ such that $A$ is separable and uncorrelated from its environment.

We call *process* of a system $A$ from state $A_1$ to state $A_2$ a time evolution $(AB)_1 \to (AB)_2$ of the isolated system $AB$ from $(AB)_1$ (with $A$ in state $A_1$) to $(AB)_2$ (with $A$ in state $A_2$), where $B$ is the environment of $A$. A process of $A$ is *reversible* if the isolated system $AB$ can undergo a time evolution $(AB)_2 \to (AB)_1$ which restores it in its initial state $(AB)_1$ and is called the *reverse* of $(AB)_1 \to (AB)_2$. A process of a system $A$ is called a *cycle* for $A$ if the final state $A_2$ coincides with the initial state $A_1$. A cycle for $A$ is not necessarily a cycle for $AB$.

An *elementary mechanical system* is a system such that the only admissible change of state for it is a space translation in a uniform external force field. An example is a particle which can only change its height in a uniform external gravitational field. A process of a system $A$ from state $A_1$ to $A_2$, such that both at $A_1$ and at $A_2$ system $A$ is separable from its environment, is called a *weight process* for $A$ if the only net effect of the process in the environment of $A$ is the change of state of an elementary mechanical system. We call *work done by a system A* in a weight process for $A$ from $A_1$ to $A_2$, denoted by $W_{12}^{A\to}$, the corresponding change in potential energy of the elementary mechanical system. The opposite of $W_{12}^{A\to}$, denoted by $W_{12}^{A\leftarrow}$, is called *work received by A* in the process.

An *equilibrium* state of a system is a state such that the system is separable and the state does not vary with time and can be reproduced while the system is isolated. An *equilibrium* state of a closed system $A$ in which $A$ is uncorrelated from its environment $B$ is called a *stable equilibrium state* if it cannot be changed to a different state in which $A$ is separable and uncorrelated from its environment by a process such that neither the regions of space where the constituents of $A$ are contained nor the state of the environment $B$ of $A$ have net changes. Two systems, $A$ and $B$, are in *mutual stable equilibrium* if the composite system $AB$ (*i.e.*, the union of both systems) is in a stable equilibrium state.

## 3. Definition of energy for a closed system

**Weight polygonal and work in a weight polygonal.** Consider an ordered set of $n + 2$ states of a closed system $A$, $(A_1, A_{i1}, A_{i2}, \ldots, A_{in-1}, A_{in}, A_2)$, such that in each of these states $A$ is separable from its environment. If $n + 1$ weight processes exist, which interconnect $A_1$ and $A_{i1}$, $A_{i1}$ and $A_{i2}$, $\ldots$, $A_{in-1}$ and $A_{in}$, $A_n$ and $A_2$, regardless of the direction of each process, we say that $A_1$ and $A_2$ can be interconnected by a weight polygonal. For instance, if weight processes $A_1 \to A_3$ and $A_2 \to A_3$ exist for $A$, we say that $A_1 \to A_3 \leftarrow A_2$ is a weight polygonal for $A$ from $A_1$ to $A_2$. We call *work done by A in a weight polygonal from $A_1$ to $A_2$* the sum of the works done by $A$ in the weight processes with direction from $A_1$ to $A_2$ and the opposites of the works done by $A$ in the weight processes with direction from $A_2$ to $A_1$ [15]. The work done by $A$ in a weight polygonal from $A_1$ to $A_2$ will be denoted by $W_{12}^{A \to \text{wpol}}$; its opposite will be called *work received by A in a weight polygonal from $A_1$ to $A_2$* and will be denoted by $W_{12}^{A \leftarrow \text{wpol}}$. Clearly, for a given weight polygonal, $W_{12}^{A \leftarrow \text{wpol}} = -W_{12}^{A \to \text{wpol}} = W_{21}^{A \to \text{wpol}}$.

In the example of weight polygonal $A_1 \to A_3 \leftarrow A_2$ sketched in Figure 1, one has that

$$W_{12}^{A \to \text{wpol}} = W_{13}^{A \to} - W_{23}^{A \to} \ . \tag{9}$$

**Figure 1.** Sketch of a weight polygonal $A_1 \to A_3 \leftarrow A_2$; the arrows represent weight processes for $A$.

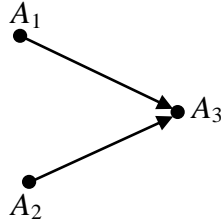

**Assumption 1**. *Every pair of states ($A_1$, $A_2$) of a closed system A, such that A is separable from its environment in both states, can be interconnected by means of a weight polygonal for A.*

**Postulate 1**. *The works done by a system in any two weight polygonals between the same initial and final states are identical.*

*Comment.* In Ref. [15] it is proved that, in sets of states where sufficient conditions of interconnectability by weight processes hold, Postulate 1 can be proved as a consequence of the traditional form of the First Law, which concerns *weight processes* (or adiabatic processes).

**Definition of energy for a closed system and proof that it is a property**. Let ($A_1$, $A_2$) be any pair of states of a closed system $A$, such that $A$ is separable from its environment in both states. We call the *energy difference* between states $A_2$ and $A_1$ the work received by $A$ in any weight polygonal from $A_1$ to $A_2$ expressed as

$$E_2^A - E_1^A = -W_{12}^{A \to \text{wpol}} = W_{12}^{A \leftarrow \text{wpol}} \ . \tag{10}$$

Assumption 1 and Postulate 1 yield the following consequences:
(a) the energy difference between two states $A_2$ and $A_1$ depends only on the states $A_1$ and $A_2$;
(b) (*additivity of energy differences*) consider a pair of states $(AB)_1$ and $(AB)_2$ of a composite system $AB$, where both $A$ and $B$ are closed and denote by $A_1$, $B_1$ and $A_2$, $B_2$ the corresponding states of $AB$; then, if $A$, $B$ and $AB$ are separable from their environment in the states considered,

$$E_2^{AB} - E_1^{AB} = E_2^A - E_1^A + E_2^B - E_1^B \ ; \tag{11}$$

(c) (*energy is a property*) let $A_0$ be a reference state of a system $A$, in which $A$ is separable from its environment, to which we assign an arbitrarily chosen value of energy $E_0^A$; the value of the energy of $A$ in any other state $A_1$ in which $A$ is separable from its environment is determined uniquely by

$$E_1^A = E_0^A + W_{01}^{A \leftarrow \text{wpol}} \quad , \tag{12}$$

where $W_{01}^{A \leftarrow \text{wpol}}$ is the work received by $A$ in any weight polygonal for $A$ from $A_0$ to $A_1$;

(d) energy is defined for every set of states of any closed system $A$ in which $A$ is separable from its environment and Assumption 1 holds.

Simple proofs of these consequences can be found in Section 5 of Ref. [15] and will not be repeated here.

*Comment*. Since the energy of $A$ is defined only when $A$ is separable from its environment, in the following we will consider as understood that $A$ is separable from its environment in every state in which the energy of $A$ is defined.

## 4. Definition of entropy for a closed system

**Postulate 2**. *Among all the states of a closed system $A$ such that the constituents of $A$ are contained in a given set of regions of space, there is a stable equilibrium state for every value of the energy $E^A$.*

**Assumption 2.** *Starting from any state in which the system is separable from its environment, a closed system $A$ can be changed to a stable equilibrium state with the same energy by means of a zero work weight process for $A$ in which the regions of space occupied by the constituents of $A$ have no net changes.*

**Lemma 1. Uniqueness of the stable equilibrium state for a given value of the energy.** There can be no pair of different stable equilibrium states of a closed system $A$ with identical regions of space occupied by the constituents of $A$ and the same value of the energy $E^A$.

**Proof.** Since $A$ is closed and in any stable equilibrium state it is separable and uncorrelated from its environment, if two such states existed, by *Assumption* 2 the system could be changed from one to the other by means of a zero-work weight process, with no change of the regions of space occupied by the constituents of $A$ and no change of the state of the environment of $A$. Therefore, neither would satisfy the definition of stable equilibrium state.

**Postulate 3**. *There exist systems, called normal systems, whose energy has no upper bound. Starting from any state in which the system is separable from its environment, a normal system $A$ can be changed to a non-equilibrium state with arbitrarily higher energy by means of a weight process for $A$ in which the regions of space occupied by the constituents of $A$ have no net changes.*

*Comments*. Exceptions to this hypothesis are *special systems*, such as *spin* systems or systems that can access only a finite number of energy levels, for which the admissible values of the energy are bounded within a finite range. The additivity of energy implies that the union of two or more normal systems, each separable from its environment, is a normal system to which Postulate 3 applies. In traditional treatments of thermodynamics, Postulate 3 is *not stated explicitly* but is used, for example, when one states that any amount of work can be transferred to a thermal reservoir by a stirrer.

**Theorem 1. Impossibility of a Perpetual Motion Machine of the Second Kind (PMM2).** If a normal system *A* is in a stable equilibrium state, it is impossible to lower its energy by means of a weight process for *A* in which the regions of space occupied by the constituents of *A* have no net change.

**Proof.** (See sketch in Figure 2). Suppose that, starting from a stable equilibrium state $A_{se}$ of *A*, by means of a weight process $\Pi_1$ with positive work $W^{A\rightarrow} = W > 0$, the energy of *A* is lowered and the regions of space occupied by the constituents of *A* have no net change. On account of Postulate 3, it would then be possible to perform a weight process $\Pi_2$ for *A* in which the regions of space occupied by the constituents of *A* have no net change, the weight *M* is restored to its initial state so that the positive amount of energy $W^{A\leftarrow} = W > 0$ is supplied back to *A*, and the final state of *A* is a nonequilibrium state, namely, a state clearly different from $A_{se}$. Thus, the composite zero-work weight process ($\Pi_1$, $\Pi_2$) would violate the definition of stable equilibrium state.

**Figure 2.** Illustration of the proof of Theorem 1.

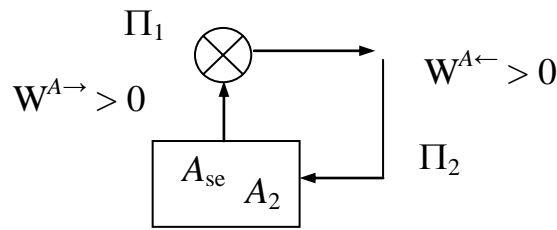

**Systems in mutual stable equilibrium**. We say that two systems *A* and *B*, each in a stable equilibrium state, are in mutual stable equilibrium if the composite system *AB* is in a stable equilibrium state.

**Assumption 3**. *There exist systems, called thermal reservoirs, which fulfill the following conditions:*
*(a) the regions of space occupied by the constituents are fixed;*
*(b) if R is a thermal reservoir in an arbitrary stable equilibrium state $R_1$ and $R^d$ is an identical copy of R also in arbitrary stable equilibrium state $R^d_2$, not necessarily equal to $R_1$, then R and $R^d$ are in mutual stable equilibrium.*

*Comment*. Every normal single-constituent system without internal boundaries and applied external fields, and with a number of particles of the order of one mole (so that the simple system approximation as defined in page 263 of Ref. [14] applies), when restricted to a fixed region of space of appropriate volume and to the range of energy values corresponding to the so-called triple-point stable equilibrium states, is an excellent approximation of a thermal reservoir.

Indeed, for a system of this kind, when three different phases (such as, solid, liquid and vapor) are present, two stable equilibrium states with different energy values have, with an extremely high approximation, the same temperature (here not yet defined) and, thus, fulfill the condition for the mutual stable equilibrium of the system and a copy thereof.

The existence of thermal reservoirs has been implicitly assumed in almost every traditional treatment of thermodynamics. It has been a basic assumption in treatments of the physics of open quantum systems [20], fluctuation theory [21], quantum measurement [22], and thermodynamics in the quantum regime [23].

**Reference thermal reservoir**. A thermal reservoir chosen once and for all is called a *reference thermal reservoir*. To fix ideas, we choose water as the constituent of our reference thermal reservoir, i.e., sufficient amounts of ice, liquid water, and water vapor at triple point conditions.

**Standard weight process**. Given a pair of states ($A_1$, $A_2$) of a closed system $A$ and a thermal reservoir $R$, we call standard weight process for $AR$ from $A_1$ to $A_2$ a weight process for the composite system $AR$ in which the end states of $R$ are stable equilibrium states. We denote by $(A_1 R_1 \to A_2 R_2)^{sw}$ a standard weight process for $AR$ from $A_1$ to $A_2$ and by $(\Delta E^R)^{sw}_{A_1 A_2}$ the corresponding energy change of the thermal reservoir $R$.

**Assumption 4**. *Every pair of states ($A_1$, $A_2$) of a closed system $A$, such that $A$ is separable and uncorrelated from its environment in both states, can be interconnected by a reversible standard weight process for $AR$, where $R$ is an arbitrarily chosen thermal reservoir initially in an arbitrarily chosen stable equilibrium state.*

**Theorem 2**. For a given closed system $A$ and a given thermal reservoir $R$, among all the standard weight processes for $AR$ between a given pair of states ($A_1$, $A_2$) of $A$ in which $A$ is separable and uncorrelated from its environment, the energy change $(\Delta E^R)^{sw}_{A_1 A_2}$ of the thermal reservoir $R$ has a lower bound which is reached if and only if the process is reversible.

**Proof**. Let $\Pi_{AR}$ denote a standard weight process for $AR$ from $A_1$ to $A_2$, and $\Pi_{AR\,\text{rev}}$ a reversible one; the energy changes of $R$ in processes $\Pi_{AR}$ and $\Pi_{AR\,\text{rev}}$ are, respectively, $(\Delta E^R)^{sw}_{A_1 A_2}$ and $(\Delta E^R)^{sw\,\text{rev}}_{A_1 A_2}$. With the help of Figure 3, we will prove that regardless of the initial state of $R$:

(a) $(\Delta E^R)^{sw\,\text{rev}}_{A_1 A_2} \leq (\Delta E^R)^{sw}_{A_1 A_2}$;

(b) if also $\Pi_{AR}$ is reversible, then $(\Delta E^R)^{sw}_{A_1 A_2} = (\Delta E^R)^{sw\,\text{rev}}_{A_1 A_2}$;

(c) if $(\Delta E^R)^{sw}_{A_1 A_2} = (\Delta E^R)^{sw\,\text{rev}}_{A_1 A_2}$, then also $\Pi_{AR}$ is reversible.

**Figure 3.** Illustration of the proof of Theorem 2.

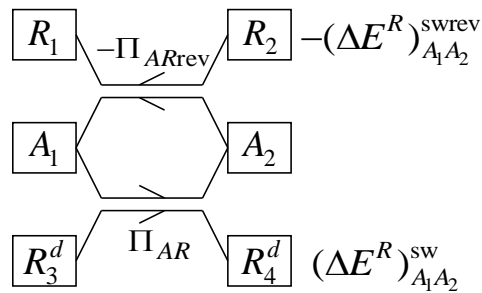

**Proof of (a).** Let us denote by $R_1$ and $R_2$ the initial and the final states of $R$ in process $\Pi_{AR\,\text{rev}}$. Let us denote by $R^d$ the duplicate of $R$ which is employed in process $\Pi_{AR}$, and by $R_3^d$ and $R_4^d$ the initial and the final states of $R^d$ in this process. Let us suppose *ab absurdo* that $\left(\Delta E^R\right)_{A_1A_2}^{\text{sw rev}} > \left(\Delta E^R\right)_{A_1A_2}^{\text{sw}}$, and consider the composite process $(-\Pi_{AR\,\text{rev}}, \Pi_{AR})$, where $-\Pi_{AR\,\text{rev}}$ is a reverse of $\Pi_{AR\,\text{rev}}$. This process would be a weight process for $RR^d$ in which, starting from the stable equilibrium state $R_2R_3^d$, the energy of $RR^d$ is lowered and the regions of space occupied by the constituents of $RR^d$ have no net changes, in contradiction to Theorem 1. Therefore, $\left(\Delta E^R\right)_{A_1A_2}^{\text{sw rev}} \leq \left(\Delta E^R\right)_{A_1A_2}^{\text{sw}}$.

**Proof of (b).** If $\Pi_{AR}$ is reversible too, then, in addition to $\left(\Delta E^R\right)_{A_1A_2}^{\text{sw rev}} \leq \left(\Delta E^R\right)_{A_1A_2}^{\text{sw}}$, the relation $\left(\Delta E^R\right)_{A_1A_2}^{\text{sw}} \leq \left(\Delta E^R\right)_{A_1A_2}^{\text{sw rev}}$ must hold too. Otherwise, the composite process $(\Pi_{AR\,\text{rev}}, -\Pi_{AR})$ would be a weight process for $RR^d$ in which, starting from the stable equilibrium state $R_1R_4^d$, the energy of $RR^d$ is lowered and the regions of space occupied by the constituents of $RR^d$ have no net changes, in contradiction to Theorem 1. Therefore, $\left(\Delta E^R\right)_{A_1A_2}^{\text{sw}} = \left(\Delta E^R\right)_{A_1A_2}^{\text{sw rev}}$.

**Proof of (c).** Let $\Pi_{AR}$ be a standard weight process for $AR$, from $A_1$ to $A_2$, such that $\left(\Delta E^R\right)_{A_1A_2}^{\text{sw}} = \left(\Delta E^R\right)_{A_1A_2}^{\text{sw rev}}$, and let $R_1$ be the initial state of $R$ in this process. Let $\Pi_{AR\,\text{rev}}$ be a reversible standard weight process for $AR$, from $A_1$ to $A_2$, with the same initial state $R_1$ of $R$. Thus, $R_3^d$ coincides with $R_1$ and $R_4^d$ coincides with $R_2$. The composite process $(\Pi_{AR}, -\Pi_{AR\,\text{rev}})$ is a cycle for the isolated system $ARB$, where $B$ is the environment of $AR$. As a consequence, $\Pi_{AR}$ is reversible, because it is a part of a cycle of the isolated system $ARB$.

**Theorem 3.** Let $R'$ and $R''$ be any two thermal reservoirs and consider the energy changes, $\left(\Delta E^{R'}\right)_{A_1A_2}^{\text{swrev}}$ and $\left(\Delta E^{R''}\right)_{A_1A_2}^{\text{swrev}}$ respectively, in the reversible standard weight processes $\Pi_{AR'} = \left(A_1R_1' \to A_2R_2'\right)^{\text{swrev}}$ and $\Pi_{AR''} = \left(A_1R_1'' \to A_2R_2''\right)^{\text{swrev}}$, where $(A_1, A_2)$ is an arbitrarily chosen pair of states of any closed system $A$, such that $A$ is separable and uncorrelated from its environment in both states. Then the ratio $\left(\Delta E^{R'}\right)_{A_1A_2}^{\text{swrev}} \Big/ \left(\Delta E^{R''}\right)_{A_1A_2}^{\text{swrev}}$

(a) is positive;
(b) depends only on $R'$ and $R''$, i.e., it is independent of (*i*) the initial stable equilibrium states of $R'$ and $R''$, (*ii*) the choice of system $A$, and (*iii*) the choice of states $A_1$ and $A_2$.

**Figure 4.** Illustration of the proof of Theorem 3, part (a).

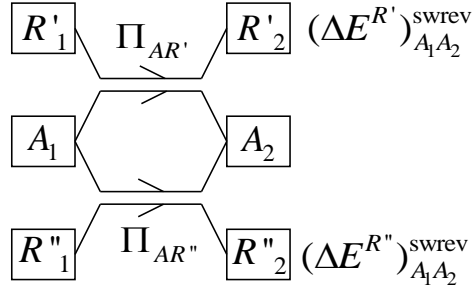

**Proof of (a).** With the help of Figure 4, let us suppose that $\left(\Delta E^{R'}\right)^{\text{swrev}}_{A_1 A_2} < 0$. Then, $\left(\Delta E^{R''}\right)^{\text{swrev}}_{A_1 A_2}$ cannot be zero. In fact, in that case the composite process $(\Pi_{AR'}, -\Pi_{AR''})$, which is a cycle for $A$, would be a weight process for $R'$ in which, starting from the stable equilibrium state $R'_1$, the energy of $R'$ is lowered and the regions of space occupied by the constituents of $R'$ have no net changes, in contradiction to Theorem 1. Moreover, $\left(\Delta E^{R''}\right)^{\text{swrev}}_{A_1 A_2}$ cannot be positive. In fact, if it were positive, the work performed by $R'R''$ as a result of the overall weight process $(\Pi_{AR'}, -\Pi_{AR''})$ for $R'R''$ would be

$$W^{R'R'' \to} = -\left(\Delta E^{R'}\right)^{\text{swrev}}_{A_1 A_2} + \left(\Delta E^{R''}\right)^{\text{swrev}}_{A_1 A_2} \quad , \tag{13}$$

where both terms are positive. After the process $(\Pi_{AR'}, -\Pi_{AR''})$, on account of Postulate 3 and *Assumption* 2, one could perform a weight process $\Pi_{R''}$ for $R''$ in which a positive amount of energy equal to $\left(\Delta E^{R''}\right)^{\text{swrev}}_{A_1 A_2}$ is given back to $R''$ and the latter is restored to its initial stable equilibrium state. As a result, the composite process $(\Pi_{AR'}, -\Pi_{AR''}, \Pi_{R''})$ would be a weight process for $R'$ in which, starting from the stable equilibrium state $R'_1$, the energy of $R'$ is lowered and the regions of space occupied by the constituents of $R'$ have no net changes, in contradiction to Theorem 1. Therefore, the condition $\left(\Delta E^{R'}\right)^{\text{swrev}}_{A_1 A_2} < 0$ implies $\left(\Delta E^{R''}\right)^{\text{swrev}}_{A_1 A_2} < 0$.

Let us suppose that $\left(\Delta E^{R'}\right)^{\text{swrev}}_{A_1 A_2} > 0$. Then, for process $-\Pi_{AR'}$ one has that $\left(\Delta E^{R'}\right)^{\text{swrev}}_{A_2 A_1} < 0$. By repeating the previous argument, one proves that for process $-\Pi_{AR''}$ one has that $\left(\Delta E^{R''}\right)^{\text{swrev}}_{A_2 A_1} < 0$. Therefore, for process $\Pi_{AR''}$ one has that $\left(\Delta E^{R''}\right)^{\text{swrev}}_{A_1 A_2} > 0$.

**Proof of (b).** Choose a pair of states $(A_1, A_2)$ of a closed system $A$ such that $A$ is separable and uncorrelated from its environment and consider the reversible standard weight process $\Pi_{AR'} = \left(A_1 R'_1 \to A_2 R'_2\right)^{\text{swrev}}$ for $AR'$ with $R'$ initially in state $R'_1$ and the reversible standard weight process $\Pi_{AR''} = \left(A_1 R''_1 \to A_2 R''_2\right)^{\text{swrev}}$ for $AR''$ with $R''$ initially in state $R''_1$. Then choose a pair of states $(A'_1, A'_2)$ of another closed system $A'$ such that $A'$ is separable and uncorrelated from its environment and consider the reversible standard weight process $\Pi_{A'R'} = \left(A'_1 R'_1 \to A'_2 R'_2\right)^{\text{swrev}}$ for $A'R'$ with $R'$ initially in state $R'_1$ and the reversible standard weight process $\Pi_{A'R''} = \left(A'_1 R''_1 \to A'_2 R''_2\right)^{\text{swrev}}$ for $A'R''$ with $R''$ initially in state $R''_1$.

With the help of Figure 5, we will prove that the changes in energy of the reservoirs in these processes obey the relation

$$\frac{\left(\Delta E^{R'}\right)^{\text{swrev}}_{A_1A_2}}{\left(\Delta E^{R''}\right)^{\text{swrev}}_{A_1A_2}} = \frac{\left(\Delta E^{R'}\right)^{\text{swrev}}_{A'_1A'_2}}{\left(\Delta E^{R''}\right)^{\text{swrev}}_{A'_1A'_2}} \quad . \tag{14}$$

Let us assume: $\left(\Delta E^{R'}\right)^{\text{swrev}}_{A_1A_2} > 0$ and $\left(\Delta E^{R'}\right)^{\text{swrev}}_{A'_1A'_2} > 0$, which implies, on account of part (a) of the proof, that $\left(\Delta E^{R''}\right)^{\text{swrev}}_{A_1A_2} > 0$ and $\left(\Delta E^{R''}\right)^{\text{swrev}}_{A'_1A'_2} > 0$. This is not a restriction, because it is possible to reverse the processes under consideration.

Now, as is well known, any real number can be approximated with arbitrarily high accuracy by a rational number. Therefore, we will assume that the energy changes $\left(\Delta E^{R'}\right)^{\text{swrev}}_{A_1A_2}$ and $\left(\Delta E^{R'}\right)^{\text{swrev}}_{A'_1A'_2}$ are rational numbers so that whatever is the value of their ratio, there exist two positive integers $m$ and $n$ such that $\left(\Delta E^{R'}\right)^{\text{swrev}}_{A_1A_2} / \left(\Delta E^{R'}\right)^{\text{swrev}}_{A'_1A'_2} = n/m$, i.e.,

$$m\left(\Delta E^{R'}\right)^{\text{swrev}}_{A_1A_2} = n\left(\Delta E^{R'}\right)^{\text{swrev}}_{A'_1A'_2} \quad . \tag{15}$$

**Figure 5.** Illustration of the proof of Theorem 3, part (b).

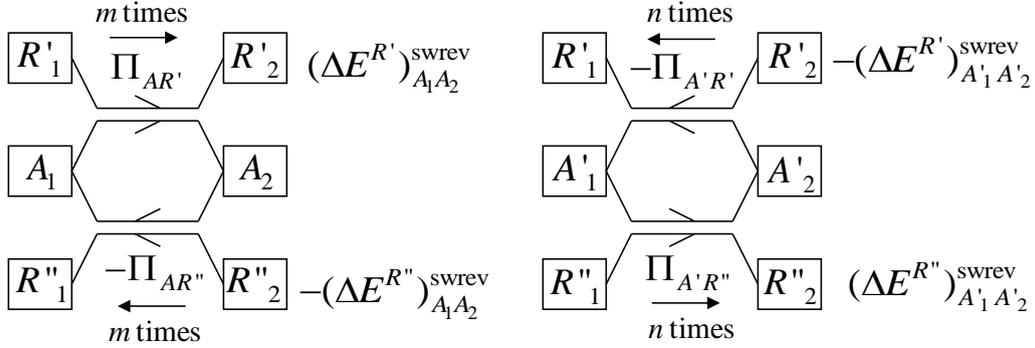

As sketched in Figure 5, let us consider the composite processes $\Pi_A$ and $\Pi_{A'}$ defined below. $\Pi_A$ is the following composite weight process for the composite system $AR'R''$: starting from the initial state $R'_1$ of $R'$ and $R''_2$ of $R''$, system $A$ is brought from $A_1$ to $A_2$ by a reversible standard weight process for $AR'$ and then from $A_2$ to $A_1$ by a reversible standard weight process for $AR''$; whatever the new states of $R'$ and $R''$ are, again system $A$ is brought from $A_1$ to $A_2$ by a reversible standard weight process for $AR'$ and back to $A_1$ by a reversible standard weight process for $AR''$ until the cycle for $A$ is repeated $m$ times. Similarly, $\Pi_{A'}$ is a composite weight process for the composite system $A'R'R''$ whereby starting from the end states of $R'$ and $R''$ reached by process $\Pi_A$, system $A'$ is brought from $A'_1$ to $A'_2$ by a reversible standard weight process for $A'R''$; then from $A'_2$ to $A'_1$ by a reversible standard weight process for $A'R'$; and so on until the cycle for $A'$ is repeated $n$ times.

Clearly, the whole composite process ($\Pi_A$, $\Pi_{A'}$) is a cycle for $AA'$. Moreover, it is a cycle also for $R'$. In fact, on account of Theorem 2, the energy change of $R'$ in each process $\Pi_{AR'}$ is equal to $\left(\Delta E^{R'}\right)^{\text{swrev}}_{A_1A_2}$, regardless of its initial state, and in each process $-\Pi_{A'R'}$ is equal to $-\left(\Delta E^{R'}\right)^{\text{swrev}}_{A'_1A'_2}$. Therefore, the energy change of $R'$ in the whole composite process ($\Pi_A$, $\Pi_{A'}$) is

$m\left(\Delta E^{R'}\right)^{\text{swrev}}_{A_1A_2} - n\left(\Delta E^{R'}\right)^{\text{swrev}}_{A'_1A'_2}$ and equals zero on account of Eq. (15). As a result, after $(\Pi_A, \Pi_{A'})$, reservoir $R'$ has been restored to its initial state so that $(\Pi_A, \Pi_{A'})$ is a reversible weight process for $R''$.

Again, on account of Theorem 2, the overall energy change of $R''$ in the whole composite process is $-m\left(\Delta E^{R''}\right)^{\text{swrev}}_{A_1A_2} + n\left(\Delta E^{R''}\right)^{\text{swrev}}_{A'_1A'_2}$. If this quantity were negative, Theorem 1 would be violated. If this quantity were positive, Theorem 1 would also be violated by the reverse of the process, $(-\Pi_A, -\Pi_{A'})$. Therefore, the only possibility is that $-m\left(\Delta E^{R''}\right)^{\text{swrev}}_{A_1A_2} + n\left(\Delta E^{R''}\right)^{\text{swrev}}_{A'_1A'_2} = 0$, i.e.,

$$m\left(\Delta E^{R''}\right)^{\text{swrev}}_{A_1A_2} = n\left(\Delta E^{R''}\right)^{\text{swrev}}_{A'_1A'_2} \quad . \tag{16}$$

Finally, taking the ratio of Eqs. (15) and (16), we obtain Eq. (14) which is our conclusion.

**Temperature of a thermal reservoir**. Let $R$ be a given thermal reservoir and $R^0$ a reference thermal reservoir. Select an arbitrary pair of states $(A_1, A_2)$ of a closed system $A$ such that $A$ is separable and uncorrelated from its environment in both states and consider the energy changes $\left(\Delta E^R\right)^{\text{swrev}}_{A_1A_2}$ and $\left(\Delta E^{R^0}\right)^{\text{swrev}}_{A_1A_2}$ in two reversible standard weight processes from $A_1$ to $A_2$, one for $AR$ and the other for $AR^0$, respectively. We call the *temperature* of $R$ the positive quantity

$$T^R = T^{R^0} \frac{\left(\Delta E^R\right)^{\text{swrev}}_{A_1A_2}}{\left(\Delta E^{R^0}\right)^{\text{swrev}}_{A_1A_2}} \quad , \tag{17}$$

where $T^{R^0}$ is a positive constant associated arbitrarily with the reference thermal reservoir $R^0$.

Clearly, the temperature $T^R$ of $R$ is defined only up to the arbitrary multiplicative constant $T^{R^0}$. If for $R^0$ we select a thermal reservoir consisting of ice, liquid water, and water vapor at triple-point conditions and we set $T^{R^0} = 273.16$ K, we obtain the Kelvin temperature scale.

**Corollary 1**. The ratio of the temperatures of two thermal reservoirs, $R'$ and $R''$, is independent of the choice of the reference thermal reservoir and can be measured directly as

$$\frac{T^{R'}}{T^{R''}} = \frac{\left(\Delta E^{R'}\right)^{\text{swrev}}_{A_1A_2}}{\left(\Delta E^{R''}\right)^{\text{swrev}}_{A_1A_2}} \quad , \tag{18}$$

where $\left(\Delta E^{R'}\right)^{\text{swrev}}_{A_1A_2}$ and $\left(\Delta E^{R''}\right)^{\text{swrev}}_{A_1A_2}$ are the energy changes of $R'$ and $R''$ in two reversible standard weight processes, one for $AR'$ and the other for $AR''$, which interconnect the same pair of states $(A_1, A_2)$ such that $A$ is separable and uncorrelated from its environment in both states.

**Proof**. Let $\left(\Delta E^{R^0}\right)^{\text{swrev}}_{A_1A_2}$ be the energy change of the reference thermal reservoir $R^0$ in any reversible standard weight process for $AR^0$ which interconnects the same states $(A_1, A_2)$ of $A$. From Eq. (17) we have that

$$T^{R'} = T^{R^0} \frac{\left(\Delta E^{R'}\right)^{\text{swrev}}_{A_1 A_2}}{\left(\Delta E^{R^0}\right)^{\text{swrev}}_{A_1 A_2}} \quad , \quad T^{R''} = T^{R^0} \frac{\left(\Delta E^{R''}\right)^{\text{swrev}}_{A_1 A_2}}{\left(\Delta E^{R^0}\right)^{\text{swrev}}_{A_1 A_2}} \quad , \tag{19}$$

so that the ratio $T^{R'}/T^{R''}$ is given by Eq. (18).

**Corollary 2**. Let $(A_1, A_2)$ be any pair of states of a closed system $A$ such that $A$ is separable and uncorrelated from its environment in both states and let $\left(\Delta E^R\right)^{\text{swrev}}_{A_1 A_2}$ be the energy change of a thermal reservoir $R$ with temperature $T^R$ in any reversible standard weight process for $AR$ from $A_1$ to $A_2$. Then, for the given system $A$, the ratio $\left(\Delta E^R\right)^{\text{swrev}}_{A_1 A_2}/T^R$ depends only on the pair of states $(A_1, A_2)$, i.e., it is independent of the choice of reservoir $R$ and of its initial stable equilibrium state $R_1$.

**Proof**. Let us consider two reversible standard weight processes from $A_1$ to $A_2$, one for $AR'$ and the other for $AR''$, where $R'$ is a thermal reservoir with temperature $T^{R'}$ and $R''$ is a thermal reservoir with temperature $T^{R''}$. Then, equation (18) yields

$$\frac{\left(\Delta E^{R'}\right)^{\text{swrev}}_{A_1 A_2}}{T^{R'}} = \frac{\left(\Delta E^{R''}\right)^{\text{swrev}}_{A_1 A_2}}{T^{R''}} \quad . \tag{20}$$

**Definition of entropy for a closed system and proof that it is a property**. Let $(A_1, A_2)$ be any pair of states of a closed system $A$ such that $A$ is separable and uncorrelated from its environment in both states, and let $R$ be an arbitrarily chosen thermal reservoir placed in the environment $B$ of $A$. We call the *entropy difference* between $A_2$ and $A_1$ the quantity

$$S_2^A - S_1^A = -\frac{\left(\Delta E^R\right)^{\text{swrev}}_{A_1 A_2}}{T^R} \quad , \tag{21}$$

where $\left(\Delta E^R\right)^{\text{swrev}}_{A_1 A_2}$ is the energy change of $R$ in any reversible standard weight process for $AR$ from $A_1$ to $A_2$ and $T^R$ is the temperature of $R$. On account of Corollary 2, the right hand side of Eq. (21) is determined uniquely by states $A_1$ and $A_2$.

Let $A_0$ be a reference state of $A$ in which $A$ is separable and uncorrelated from its environment and assign to $A_0$ an arbitrarily chosen value $S_0^A$ of the entropy. Then, the value of the entropy of $A$ in any other state $A_1$ of $A$ in which $A$ is separable and uncorrelated from its environment is determined uniquely by the equation

$$S_1^A = S_0^A - \frac{\left(\Delta E^R\right)^{\text{swrev}}_{A_0 A_1}}{T^R} \quad , \tag{22}$$

where $\left(\Delta E^R\right)^{\text{swrev}}_{A_0 A_1}$ is the energy change of $R$ in any reversible standard weight process for $AR$ from $A_0$ to $A_1$ and $T^R$ is the temperature of $R$.

Therefore, entropy is a property of any closed system $A$, defined in every set of states of $A$ where *Assumptions* 2 and 4 hold.

**Theorem 4**. Additivity of entropy differences. Consider the pair of states ($C_1 = A_1 B_1$, $C_2 = A_2 B_2$) of the composite system $C = AB$ such that $A$ and $B$ are closed, $A$ is separable and uncorrelated from its environment in both states $A_1$ and $A_2$, and $B$ is separable and uncorrelated from its environment in both states $B_1$ and $B_2$. Then,

$$S^{AB}_{A_2 B_2} - S^{AB}_{A_1 B_1} = S^A_2 - S^A_1 + S^B_2 - S^B_1 \quad . \tag{23}$$

**Proof**. Let us choose a thermal reservoir $R$ with temperature $T^R$ and consider the composite process ($\Pi_{AR}$, $\Pi_{BR}$) where $\Pi_{AR}$ is a reversible standard weight process for $AR$ from $A_1$ to $A_2$, while $\Pi_{BR}$ is a reversible standard weight process for $BR$ from $B_1$ to $B_2$. The composite process ($\Pi_{AR}$, $\Pi_{BR}$) is a reversible standard weight process for $CR$ from $C_1$ to $C_2$ in which the energy change of $R$ is the sum of the energy changes in the constituent processes $\Pi_{AR}$ and $\Pi_{BR}$, i.e., $\left(\Delta E^R\right)^{\text{swrev}}_{C_1 C_2} = \left(\Delta E^R\right)^{\text{swrev}}_{A_1 A_2} + \left(\Delta E^R\right)^{\text{swrev}}_{B_1 B_2}$. Therefore,

$$\frac{\left(\Delta E^R\right)^{\text{swrev}}_{C_1 C_2}}{T^R} = \frac{\left(\Delta E^R\right)^{\text{swrev}}_{A_1 A_2}}{T^R} + \frac{\left(\Delta E^R\right)^{\text{swrev}}_{B_1 B_2}}{T^R} \quad . \tag{24}$$

Equation (24) together with the definition of the entropy difference, Eq. (21), yield Eq. (23).

*Comment*. As a consequence of Theorem 4, if the values of entropy are chosen so that they are additive over the subsystems in their reference states, the entropy of a composite system is equal to the sum of the entropies of the constituent subsystems.

**Theorem 5**. Let ($A_1$, $A_2$) be any pair of states of a closed system $A$ such that $A$ is separable and uncorrelated from its environment in both states, and let $R$ be a thermal reservoir with temperature $T^R$. Furthermore, let $\Pi_{AR\text{irr}}$ be any irreversible standard weight process for $AR$ from $A_1$ to $A_2$ and let $\left(\Delta E^R\right)^{\text{swirr}}_{A_1 A_2}$ be the energy change of $R$ in this process. Then

$$-\frac{\left(\Delta E^R\right)^{\text{swirr}}_{A_1 A_2}}{T^R} < S^A_2 - S^A_1 \quad . \tag{25}$$

**Proof.** Let $\Pi_{AR\text{rev}}$ be any reversible standard weight process for $AR$ from $A_1$ to $A_2$ and let $\left(\Delta E^R\right)^{\text{swrev}}_{A_1 A_2}$ be the energy change of $R$ in this process. On account of Theorem 2,

$$\left(\Delta E^R\right)^{\text{swrev}}_{A_1 A_2} < \left(\Delta E^R\right)^{\text{swirr}}_{A_1 A_2} \quad . \tag{26}$$

Since $T^R$ is positive, from Eqs. (26) and (21) one obtains

$$-\frac{\left(\Delta E^R\right)^{\text{swirr}}_{A_1 A_2}}{T^R} < -\frac{\left(\Delta E^R\right)^{\text{swrev}}_{A_1 A_2}}{T^R} = S^A_2 - S^A_1 \quad . \tag{27}$$

**Theorem 6. (Principle of entropy nondecrease).** Let ($A_1$, $A_2$) be a pair of states of a closed system $A$ such that $A$ is separable and uncorrelated from its environment in both states, and let $(A_1 \rightarrow A_2)_w$ be any weight process for $A$ from $A_1$ to $A_2$. Then, the entropy difference $S^A_2 - S^A_1$ is equal to zero if and only if the weight process is reversible; it is strictly positive if and only if the weight process is irreversible.

**Proof**. If $(A_1 \to A_2)_w$ is reversible, then it is a special case of a reversible standard weight process for $AR$ in which the initial stable equilibrium state of $R$ does not change. Therefore, $(\Delta E^R)^{\text{swrev}}_{A_1 A_2} = 0$; and by applying the definition of entropy difference, Eq. (21), one obtains

$$S_2^A - S_1^A = -\frac{(\Delta E^R)^{\text{swrev}}_{A_1 A_2}}{T^R} = 0 \quad . \tag{28}$$

If $(A_1 \to A_2)_w$ is irreversible, then it is a special case of an irreversible standard weight process for $AR$ in which the initial stable equilibrium state of $R$ does not change. Therefore, $(\Delta E^R)^{\text{swirr}}_{A_1 A_2} = 0$, and Equation (25) yields

$$S_2^A - S_1^A > -\frac{(\Delta E^R)^{\text{swirr}}_{A_1 A_2}}{T^R} = 0 \quad . \tag{29}$$

Moreover, if a weight process $(A_1 \to A_2)_w$ for $A$ is such that $S_2^A - S_1^A = 0$, then the process must be reversible, because we just proved that for any irreversible weight process $S_2^A - S_1^A > 0$; if a weight process $(A_1 \to A_2)_w$ for $A$ is such that $S_2^A - S_1^A > 0$, then the process must be irreversible, because we just proved that for any reversible weight process $S_2^A - S_1^A = 0$.

## 5. Correspondence between the implications of Assumptions 2 and 4 and those of the *comparison hypothesis* of Lieb and Yngvason, for a collection of simple systems

In this Section, we consider a closed system $A$ which is composed of simple subsystems so that scaled copies of $A$ can be formed when its simple subsystems are in stable equilibrium. For $A$, we consider a set of states $\hat{\Gamma}$ such that, for every state, $A$ is separable and uncorrelated from its environment and energy is defined.

We prove that

1) if Postulates 2 and 3 and Assumptions 2 and 4 hold in $\hat{\Gamma}$, then the *comparison hypothesis* (i.e., every pair of states can be interconnected by a weight process for $A$) holds in $\hat{\Gamma}$ (see Theorem 6);
2) if (i) the *comparison hypothesis* holds in $\hat{\Gamma}$; (ii) the relation $\prec$ fulfils Axioms (A1) ÷ (A6) of Refs. [11, 16, 17] in the subset $\Gamma$ of the stable equilibrium states of $\hat{\Gamma}$ and Axioms N1, N2 of Ref. [17] in $\hat{\Gamma}$, so that, due to (i) and (ii) entropy is defined in $\hat{\Gamma}$ and has its well-known characteristics; (iii) every pair of stable equilibrium states of $\Gamma$ can be interconnected by a reversible standard weight process for $AR$ where $R$ is any thermal reservoir *initially in an arbitrarily chosen stable equilibrium state*; then Assumptions 2 and 4 hold in $\hat{\Gamma}$ (see Theorem 8 below).

Postulates 2 and 3, condition (ii) (for a system composed of simple subsystems) and condition (iii) can be considered as always fulfilled (the latter by any quasistatic process in which a reversible Carnot engine is employed). Therefore, one can conclude that, for a system composed of simple subsystems, the domain of validity of the definition of entropy presented here (as determined by Assumptions 2 and 4) coincides with the domain of the states in which Lieb and Yngvason prove the existence and uniqueness of an entropy function which characterizes the relation of adiabatic comparability (as determined by the *comparison hypothesis*).

**Theorem 7.** Consider a set of states $\hat{\Gamma}$ of a closed system $A$ such that in the whole set, $A$ is separable and uncorrelated from its environment, energy is defined, and Postulates 2, 3 and Assumptions 2, 4 hold. Then, every pair of states $(A_1, A_2)$ of $\hat{\Gamma}$ can be interconnected by a weight process for $A$.

**Proof**. Let $(A_1, A_2)$ be any pair of states of $\hat{\Gamma}$ and consider a reversible standard weight process $(A_1R_1 \rightarrow A_2R_2)_{sw}$ for $AR$ where $R$ is an arbitrarily chosen thermal reservoir, which exists due to Assumption 4. If the energy of $R$ in state $R_2$ coincides with that in state $R_1$, then state $R_2$ coincides with $R_1$ on account of Postulate 2 and Lemma 1 so that $(A_1R_1 \rightarrow A_2R_2)_{sw}$ is a weight process for $A$ from $A_1$ to $A_2$.

Suppose that the energy of $R$ in state $R_2$ is lower than that in state $R_1$. On account of Postulate 3 and Assumption 2, there exists a weight process for $R$ from $R_2$ to $R_1$. The composite process $(A_1R_1 \rightarrow A_2R_2)_{sw}$ followed by this process is a weight process for $A$ from $A_1$ to $A_2$, because $R$ is restored in its initial state.

Suppose that the energy of $R$ in state $R_2$ is higher than that in state $R_1$. On account of Postulate 3 and Assumption 2, there exists a weight process for $R$ from $R_1$ to $R_2$. This process, followed by the reverse of $(A_1R_1 \rightarrow A_2R_2)_{sw}$, forms a weight process for $A$ from $A_2$ to $A_1$.

*Comment*. Note that Assumptions 2 and 4 yield in $\hat{\Gamma}$ the *comparison hypothesis* and, *a fortiori*, Assumption 1. The latter is stated here as an independent assumption because it is supposed to also hold in *broader sets of states*.

**Theorem 8.** Consider a set of states $\hat{\Gamma}$ of a closed system $A$ such that, in the whole set, $A$ is separable and uncorrelated from its environment and energy is defined. Moreover,
(i) the *comparison hypothesis* holds;
(ii) the relation $\prec$ fulfils Axioms N1, N2 of Ref. [17] and Axioms (A1) ÷ (A6) of Refs. [11, 16, 17] in the subset of stable equilibrium states so that, due to (i) and (ii), entropy is defined and has its well-known characteristics;
(iii) every pair of stable equilibrium states can be interconnected by a reversible standard weight process for $AR$ where $R$ is an arbitrarily chosen thermal reservoir initially in an arbitrarily chosen stable equilibrium state.

Then Assumptions 2 and 4 hold in $\hat{\Gamma}$, namely,
(a) *starting from any state of $\hat{\Gamma}$, a system $A$ can be changed to a stable equilibrium state with the same energy by means of a zero-work weight process for $A$ in which its regions of space have no net changes*;
(b) *every pair of states $(A_1, A_2)$ of $\hat{\Gamma}$ can be interconnected by a reversible standard weight process for $AR$, where $R$ is an arbitrarily chosen thermal reservoir initially in an arbitrarily chosen stable equilibrium state*.

**Proof of (a)**. Let $A_1$ be any state of $\hat{\Gamma}$, let $A_{2se}$ be the stable equilibrium state in $\Gamma$ with the same energy and with the same regions of space occupied by the constituents of $A$, and assume $A_1$ does not coincide with $A_{2se}$ so that, by the highest entropy principle, $S_{2se}^A > S_1^A$. Due to the *comparison hypothesis*, there exists a weight process for $A$, either from $A_1$ to $A_{2se}$ or from $A_{2se}$ to $A_1$. However, since $S_{2se}^A > S_1^A$, the latter cannot exist because it would violate the principle of entropy nondecrease. Therefore, the weight process from $A_1$ to $A_{2se}$ is necessarily possible.

**Proof of (b)**. Let ($A_1$, $A_2$) be any pair of states of $A$ in $\hat{\Gamma}$ and consider the stable equilibrium states $A_{3se}$ and $A_{4se}$ in $\Gamma$ such that $A_{3se}$ has the same regions of space occupied by $A$ and the same entropy as $A_1$, while $A_{4se}$ has the same regions of space occupied by $A$ and the same entropy as $A_2$. States $A_{3se}$ and $A_{4se}$ exist because, for every spatial configuration of the system, the entropy of $A$ at equilibrium can assume all the allowed values for $A$ between the lower bound (at zero temperature) and positive infinity. On account of the *comparison hypothesis*, a weight process for $A$ which interconnects $A_1$ and $A_{3se}$ exists and is reversible due to the principle of entropy nondecrease. Similarly, a reversible weight process for $A$ exists between $A_2$ and $A_{4se}$. Finally, due to (iii), a reversible standard weight process for $AR$ between $A_{3se}$ and $A_{4se}$ exists for every choice of $R$ and of its initial stable equilibrium state. The composite of these processes is a reversible standard weight process for $AR$ between $A_1$ and $A_2$.

## 4. Conclusions

The principal methods for the definition of thermodynamic entropy have been analyzed with special reference to the most recent contributions. Then, an improvement of the treatment of the foundations of thermodynamics proposed by the present authors has been presented. In particular, the definition of energy has been extended to a set of states of a closed system $A$ where not necessarily any pair of states of $A$ can be interconnected by a weight process for $A$. Moreover, the domains of validity of the definitions of energy and of entropy have been clarified by dividing the axioms into two different groups: three *Postulates*, which are declared as having a fully general validity, and four *Assumptions*, which identify the domain of validity of the definition of energy (Assumption 1) and that of entropy (Assumptions 2, 3, 4).

It has been proven that, for a system which allows the formation of scaled copies when its subsystems are in stable equilibrium, Assumptions 2, 3, and 4 yield, in our framework, the definition of entropy in a domain which coincided with that in which Lieb and Yngvason prove, in their framework, the existence and uniqueness of an entropy function which characterizes the relation of adiadatic comparability, through the *comparison hypothesis* (every pair of states can be interconnected by a weight process for the system). As a consequence, the domain of validity of the definition of entropy presented here can be considered as an extension of the domain of validity of the definition proposed by Lieb and Yngvason to systems which do not allow the creation of scaled copies even in their stable equilibrium states.

The domain of validity of the definition of entropy presented here does not include necessarily all nonequilibrium states of any system, but, at the same time, is not necessarily restricted to stable equilibrium states nor to some particular class of systems.

## Acknowledgements

The authors are grateful to prof. Adriano M. Lezzi for helpful suggestions.